# High-frequency structure design and RF stability analysis of a 4-vane radio frequency quadrupole with pi-mode stabilizer loops

Xiaowen Zhu*, Claude Marchand, Olivier Piquet, Michel Desmons

*IRFU, CEA, Université Paris-Saclay, F-91191 Gif-sur-Yvette, France*

**Abstract:**

Compact accelerator-based neutron source facilities are garnering attention and play an important and expanding role in material and engineering sciences, as well as in neutron science education and training. Neutrons are produced by bombarding a low-energy proton beam onto a beryllium or lithium target. In such an accelerator-based neutron source, a radio frequency quadrupole (RFQ) is usually utilized to accelerate a high-intensity proton beam to a few MeV. This study mainly covers the high-frequency structure design optimizations of a 4-vane RFQ with pi-mode stabilizer loops (PISLs) and its RF stability analysis. A 176 MHz RFQ accelerator is designed to operate at a 10% duty factor and could accelerate an 80 mA proton beam from 65 keV to 2.5 MeV within a length of 5.3 m. The adoption of PISLs ensures high RF stability, eases the operation of the accelerator, and implies less stringent alignment and machining tolerances.

## 1. Introduction

Radio frequency quadrupole (RFQ) is widely utilized as an initial linear accelerator stage for both large science facilities [1,2,3,4,5], and compact accelerator-based neutron source (CANS) facilities [6,7,8,9,10]. This is because the RFQ can effectively fulfill functions of focusing, bunching, and accelerating a continuous charged particle beam simultaneously. Furthermore, 4-vane type RFQs have benefits such as lower dipole field components, higher mechanical strength, and easier cooling at high duty cycle operation. A lower operating frequency is beneficial for a high shunt impedance and modest power density design; therefore, a high-intensity 176 MHz 4-vane RFQ is studied for a new high-brilliance neutron source project proposed at CEA Paris-Saclay [11]. This 176 MHz RFQ will be designed to operate in pulsed mode with 10% duty factor, and be capable of providing an acceleration of 2.5 MeV to an input 65 keV, 80 mA proton beam in approximately 5.3 m with a tip-to-tip voltage of 80 kV.

Perturbation caused by the dipole or other modes can result in a distorted field distribution along the RFQ structure. Moreover, owing to the highly sensitive nature of the RFQ cavity, machining and assembly errors can also result in a possible dipole-mode excitation. Several design options exist to separate the frequency of the lowest-order dipole mode from that of the accelerating quadrupole mode. For example, the application of dipole stabilization rods (DSRs) inserted into the end and coupling cells increases the frequency span between the operating quadrupole mode and its nearest dipole modes [12,13,14,15,16,17]. Vane coupling rings (VCRs) are utilized to shorten the opposite vanes together, forcing them to the same potential, but with difficulties in mounting and cooling [18,19]. Another solution is to adopt pi-mode stabilizer loops (PISLs) to strongly couple adjacent RFQ quadrants, which could result in a larger mode separation from the quadrupole mode, ensuring a higher RF stability [20,21,22,23,24,25]. PISLs are chosen in our high-duty-factor RFQ design.

*Corresponding author at: IRFU, CEA, Université Paris-Saclay, F-91191 Gif-sur-Yvette, France.
*E-mail address:* zhuxw13@pku.edu.cn (X.W. Zhu)

In this study, we describe the physics design results and focus on the high-frequency structure design optimizations and the systematic analysis of RF stability in a 4-vane RFQ with PISLs. The structural design of other parts, including the tuner period, modulation perturbation, and end-cell cutbacks, are also presented. In addition, multipacting simulations are conducted for different power levels.

## 2. Proton RFQ Physics Design Results

The design parameters of the Proton RFQ are summarized in Table 1.

Table 1 Main parameters of Proton RFQ.

| Parameters | Proton RFQ |
|---|---|
| Operating frequency (MHz) | 176 |
| Beam Current (mA) | 80 |
| Input energy (keV) | 65 |
| Output energy (MeV) | 2.5 |
| Vane voltage (kV) | 80 |
| Vane length (cm) | 529.0 |
| Beam power (kW) | 200 |
| Cavity power (kW) | 211 |
| Total Power (kW) | 411 |
| Average aperture radius (mm) | 5.666 |
| Synchronous phase (degree) | -90.0 – -30.0 |
| Modulation factor | 1 – 2.382 |
| $\varepsilon_{x,y}(\mathrm{norm.rms.,entrance})\ (\mathrm{mm\cdot mrad})$ | 0.200 |
| $\varepsilon_{x}(\mathrm{norm.rms.,exit})\ (\mathrm{mm\cdot mrad})$ | 0.273 |
| $\varepsilon_{y}(\mathrm{norm.rms.,exit})\ (\mathrm{mm\cdot mrad})$ | 0.263 |
| $\varepsilon_{l}(\mathrm{norm.rms.,exit})\ (\mathrm{MeV\cdot deg})$ | 0.127 |
| Kilpatrick limit | 1.403 |
| Transmission efficiency (%) | 98.9 |

The physical design of the Proton RFQ is performed with the ParmteqM code [26] and benchmarked with the Toutatis code [27]. As illustrated in Figure 1, the multiparticle tracking simulation is performed at 80 mA with an input of $10^5$ macroparticles. The transverse distribution is a 4D waterbag, whereas the longitudinal distribution is uniform. The transmission efficiency provided by ParmteqM is 98.9%, and when tracking using Toutatis, the transmission is 99.3%. This slight difference in transmission could be explained by the different loss criteria imposed in both codes, that is, a particle travels outside the minimum aperture radius or hits a vane-tip. Despite the different criteria, the simulation results obtained from both codes agree well. The cavity length is approximately 5.3 m, average aperture radius is 5.666 mm, vane voltage is chosen to be 80 kV, and the Kilpatrick limit [28] is optimized to be 1.4 ($E_k$ = 14.016 MV/m at 176 MHz), which indicates a low risk of RF breakdown.

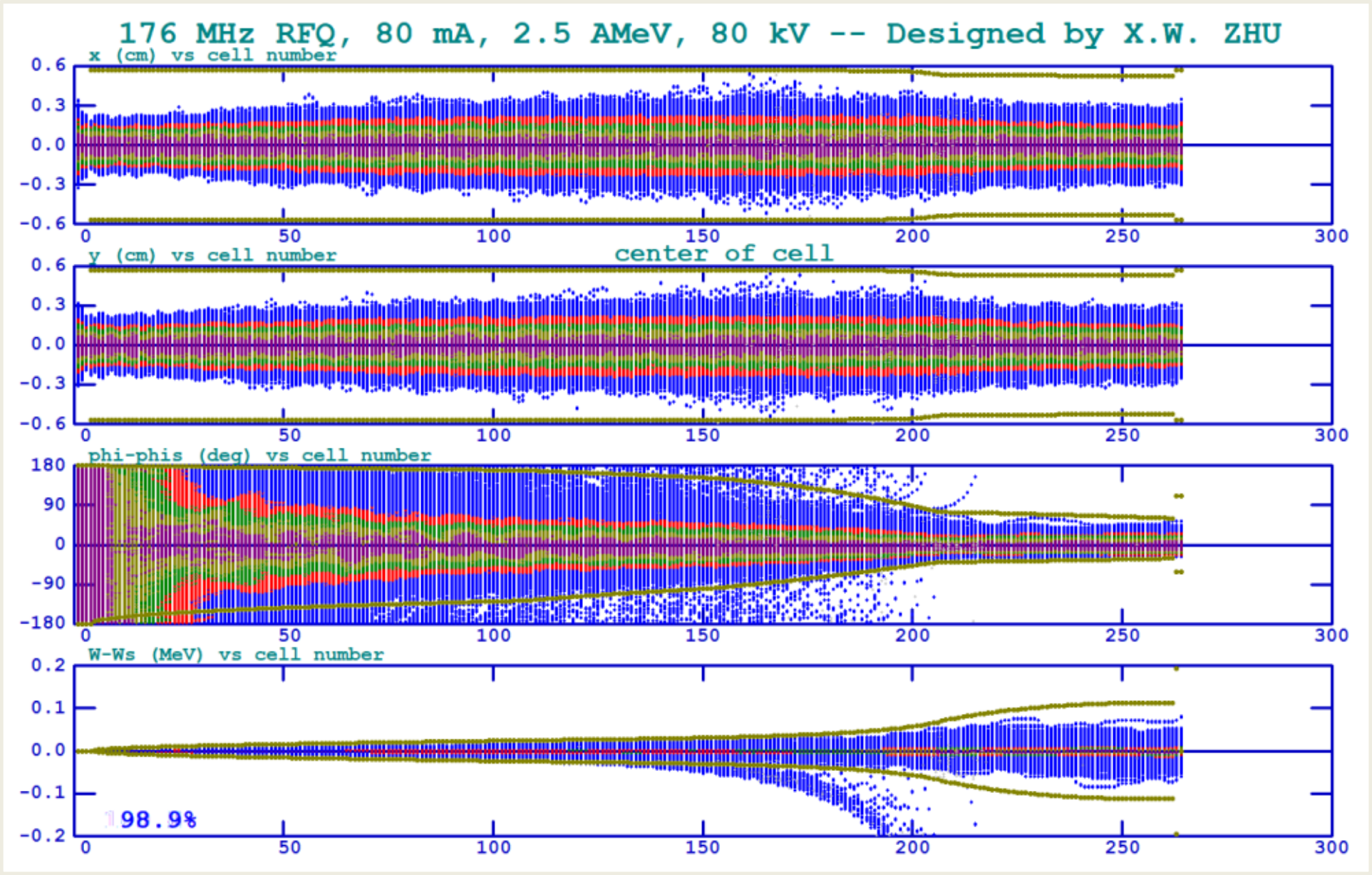

Fig.1 Particle tracking results of Proton RFQ simulated by ParmteqM at 80 mA.

Figure 2(a) illustrates the transmission efficiency curves for different input beam currents. The design indicates a transmission over 98.9% up to 80 mA, and the transmission gradually decreases along with increasing input current. Even at 160 mA, the transmission efficiency is higher than 90%. Figure 2(b) illustrates how the transmission evolves as a function of the input-normalized transverse rms emittance. When the input emittance is below 0.8 mm·mrad, the beam transmission exceeds 90%. Figure 2(c) illustrates the variation in the transmission at different input normalized power factors (P/Pnorm). When P/Pnorm is greater than 0.9, the transmission is greater than 98.5%. Figure 2(d) illustrates the changes in transmission efficiency with different input energy spreads. To achieve a high transmission over 90%, the energy spread of the input beam should be within ±10%.

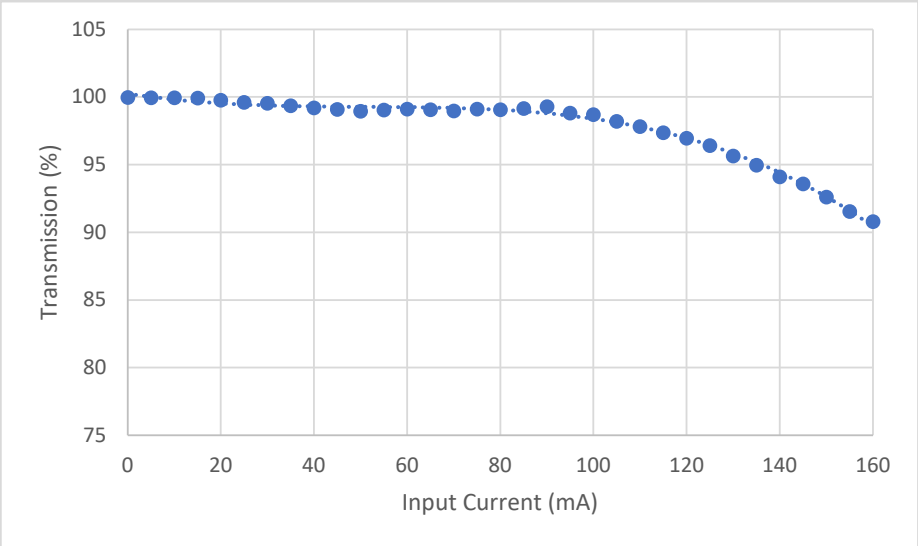

(a)

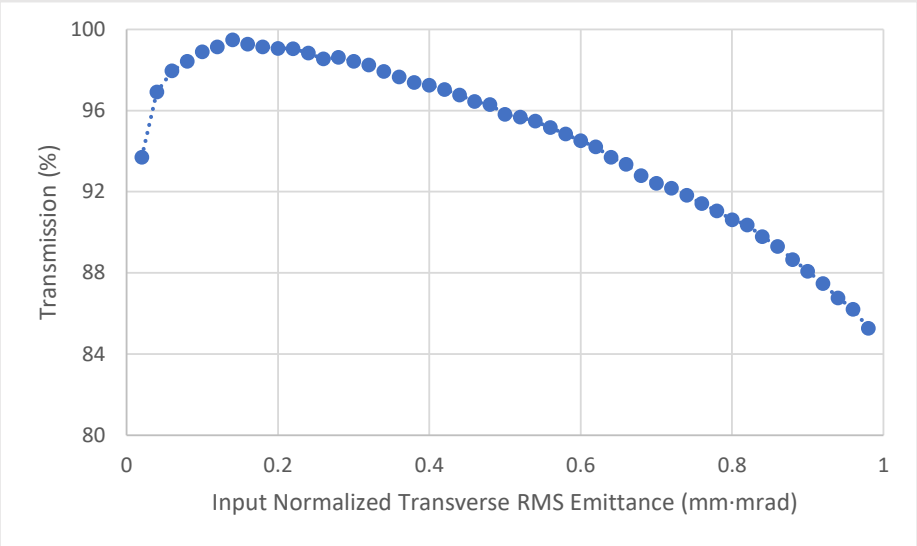

(b)

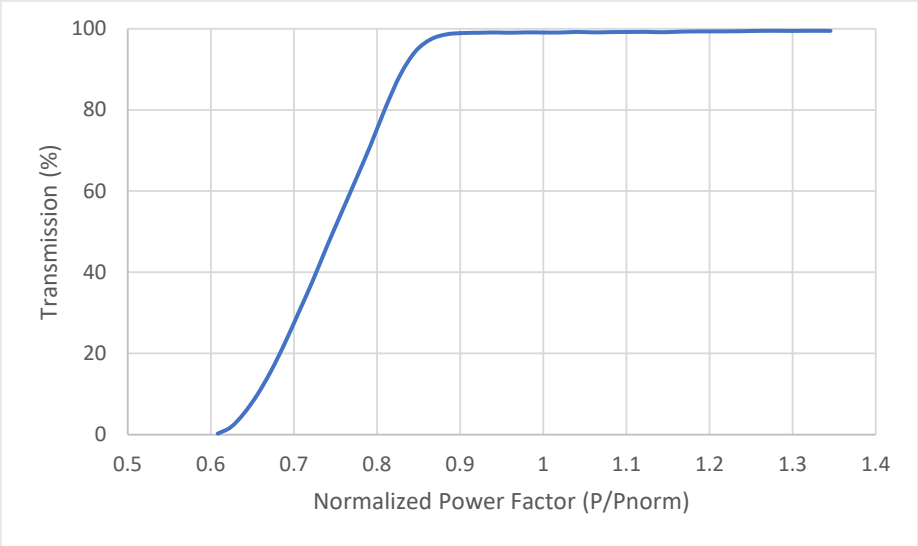

(c)

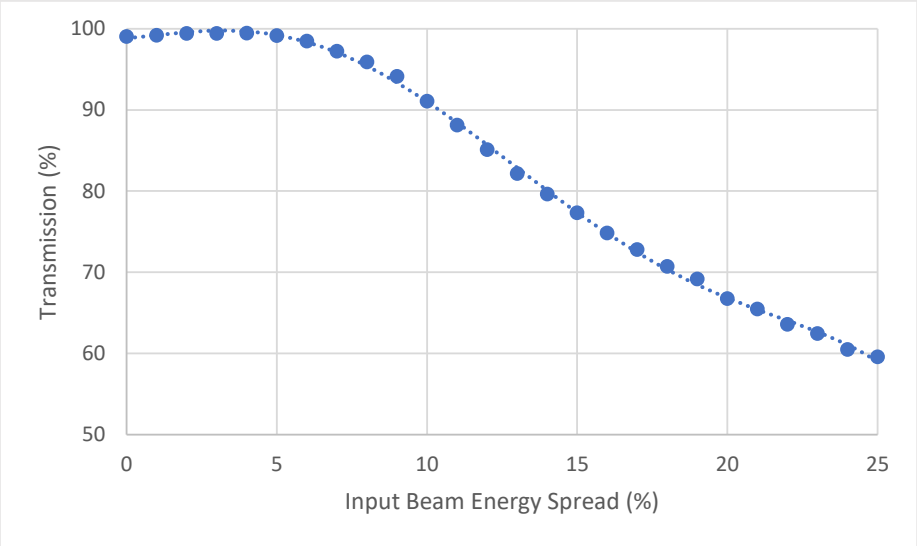

(d)

Fig.2 RFQ transmission efficiency evolution as a function of (a) input current at the designed vane voltage of 80 kV, (b) normalized transverse rms emittance, (c) input power factor (P/Pnorm), and (d) input beam energy spread.

## 3. 2D RFQ Design

The 2D cross-sectional profile is a fundamental element in the 4-vane RFQ design. A quadrilateral profile compared to an octagonal profile, is beneficial for a higher shunt impedance because of the larger area-to-perimeter ratio. It also offers the benefit of opening holes on a planar surface during machining processing.

Electromagnetic simulations of the 2D RFQ were conducted with CST MWS [29] by applying a magnetic boundary condition to the front and back ends of a thin-slice RFQ model. The geometric parameters of the cavity profiles are illustrated in Figure 3. This profile is defined with nine independent variables, and the other three are the derived parameters. To obtain a better presentation of the optimization, the structural parameters are summarized in Table 2 and then the optimization of the vane-tip radius ($r_t$) is considered as a typical example. Each time, the resonant frequency is tuned to the designed frequency of 176 MHz by adjusting the half inner width ($L_{max}$) of the RFQ cavity, and the field strength is scaled to the nominal inter-vane voltage of 80 kV.

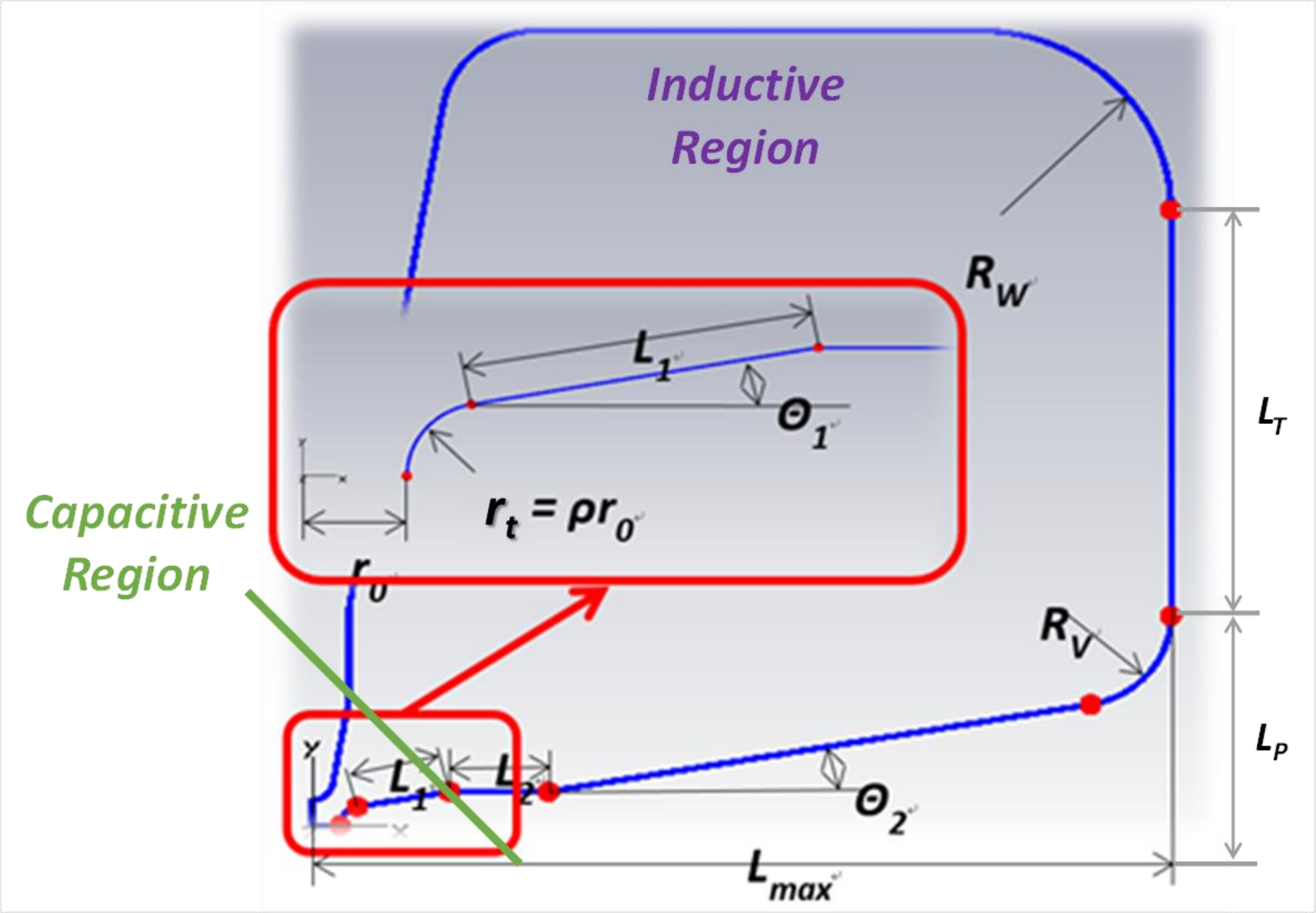

Fig.3 Cross-sectional profile of 176 MHz Proton RFQ, the dominant capacitive region is enclosed from vane-tips to the green curve, while the other region is dominant inductive.

Table 2 RF cavity profile parameters.

| Parameter | Value | Comment |
|---|---|---|
| $r_0$ (mm) | 5.666 | Average aperture radius |
| $\rho$ | 0.75 | Ratio of vane-tip radius to aperture radius |
| $L_1$ (mm) | 20 | Length of the first sloped lines from vane |
| $L_2$ (mm) | 20 | Horizontal or vertical length of the straight line |
| $\theta_1$ (degree) | 10 | First angle of the vane |
| $\theta_2$ (degree) | 10 | Second angle of the vane |
| $R_v$ (mm) | 20 | Curvature of the vane corner |
| $R_w$ (mm) | 40 | Curvature of the RFQ cavity wall |
| $L_{max}$ (mm) | 164.974 | Half of the inner width of the RFQ |
| $r_t = \rho r_0$ (mm) | 4.250 | Transverse vane-tip radius (derived parameter) |
| $L_T$ (mm) | 5.666 | Max space of tuner port (derived parameter) |
| $L_P$ (mm) | 0.75 | Min space of PISL rod (derived parameter) |

For the 2D cavity RF optimization, the following criteria are considered:

(1) Maximizing quality factor and mode separation (ΔF)

(2) Reducing peak electric and magnetic fields ($E_{max}$, $H_{max}$)

(3) Maximizing available space for tuner and PISL Rod

### 3.1 From RF Structure Design Aspect

For 2D RFQ cavity optimization, the optimization process of the tip radius ($r_t$) is a good starting example because the tip radius is a crucial parameter given by the RFQ beam dynamics under the quasistatic approximation of the RF quadrupole structure. A series of optimizations have been performed to study the influence of the vane-tip radius on cavity performance, and the results are illustrated in Figure 4. As can be observed, Figure 4(a) illustrates that both mode separation and quality factor decrease when increasing tip radius. Figure 4(b) illustrates that a larger tip radius degrades the shunt impedance, and more RF power is required to obtain the designed vane voltage. Figure 4(c) illustrates that the risk of voltage breakdown and difficulty of heat removal increases when the tip radius increases. Figure 4(d) illustrates that the available space for the tuner and PISL rod becomes smaller when a larger tip radius is applied. Therefore, a smaller vane-tip radius is more favorable for better cavity performance.

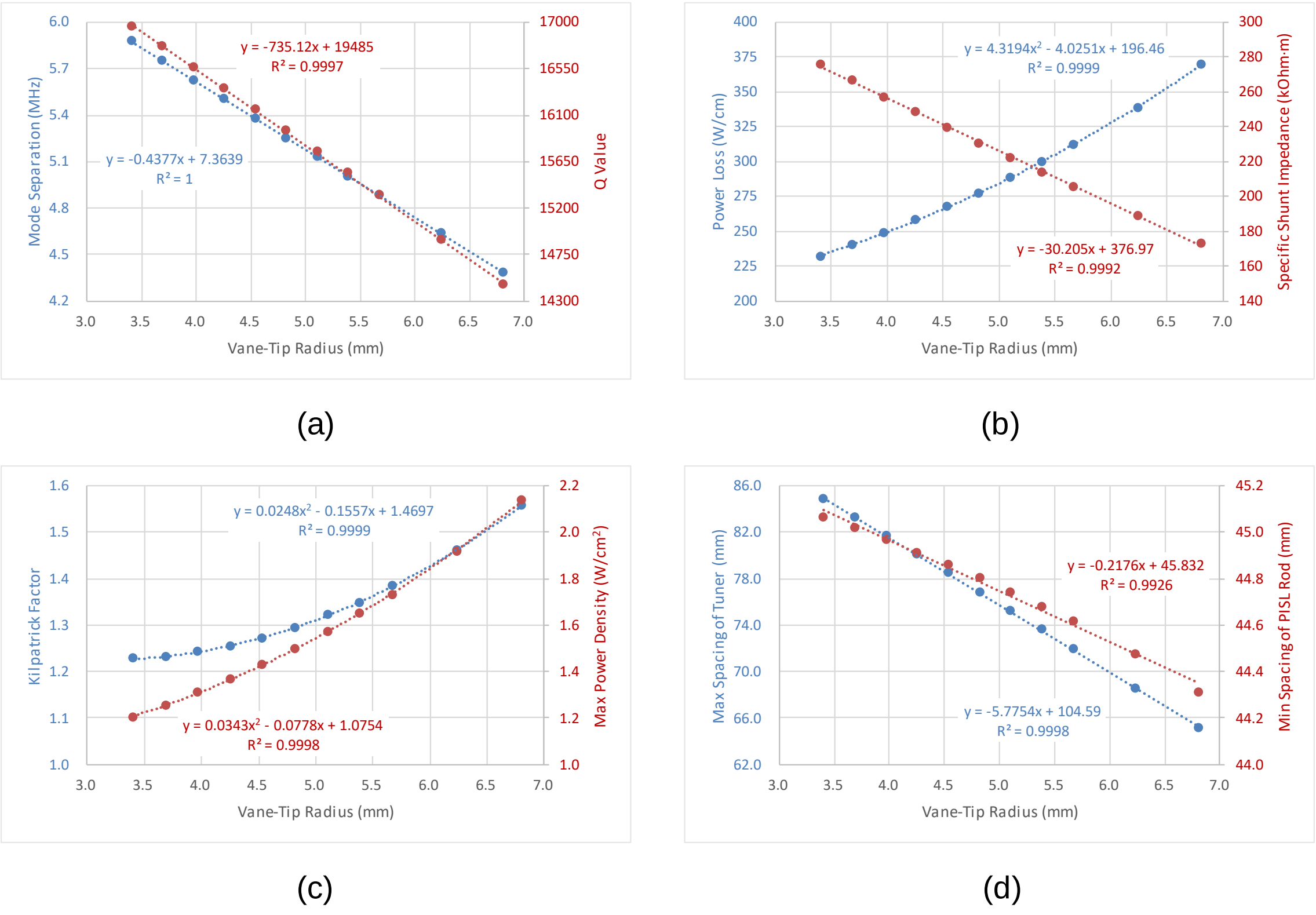

Fig.4 Performance evolutions of RF parameters with different vane-tip radius: (a) mode separation and quality factor; (b) power loss per unit length and specific shunt impedance; (c) peak surface electric field strength (normalized to Kilpatrick limit at 176 MHz) and maximum power loss density along cavity profile; (4) available space for tuner and PISL rod.

### 3.2 From Physical Design Aspect

Moreover, we must consider further optimization from the perspective of beam dynamics. The maximum peak surface electric field $E_{max}$ of each cell obtained from ParmteqM is deduced from the following expression:

$$E_{max} = k\left(\frac{r_t}{r_0}, \frac{L_{cell}}{r_0}, m\right)\frac{V}{r_0} \quad (1)$$

where k is the field enhancement factor, which can be interpolated from the tables of the VaneGeometry of ParmteqM, $r_t/r_0$ is the ratio of the vane-tip radius to the aperture radius, $L_{cell}/r_0$ denotes the ratio of a unit cell to the aperture radius, m is the modulation factor, and V represents the tip-to-tip voltage.

An investigation of the peak surface electric field at different tip radii when considering modulated vane-tips is illustrated in Figure 5. The minimum Kilpatrick factor is achieved at a tip radius of 4.250 mm.

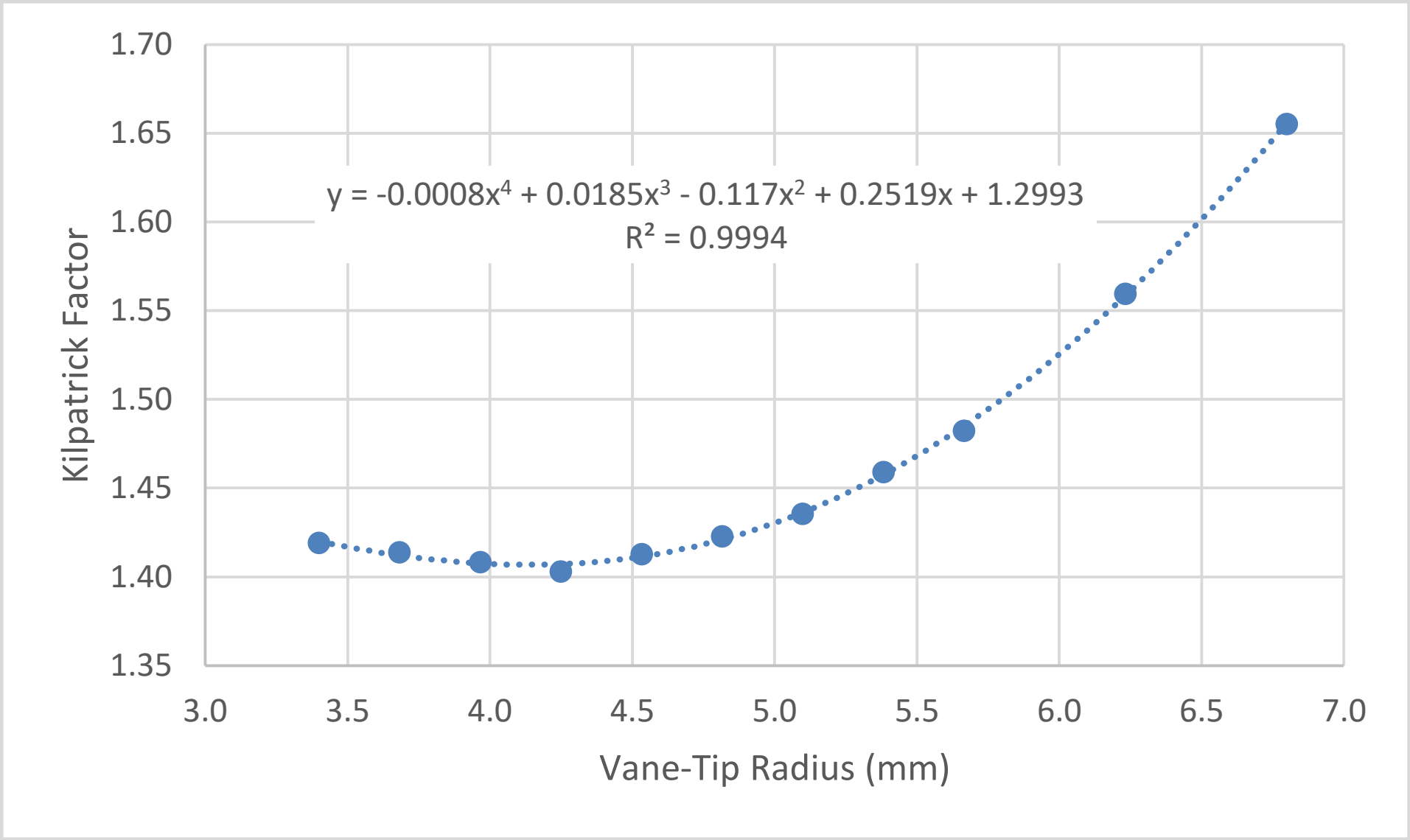

Fig.5 Evolution of Kilpatrick factor on the tip surface with modulation as a function of vane-tip radius.

In addition, the influences on transverse and longitudinal emittance growth and beam transmission are studied (Figure 6). Figure 6(a) illustrates that the transverse emittances begin to converge at a tip radius of 4.250 mm. As illustrated in Figure 6(b), although a higher transmission rate can be obtained, the longitudinal emittance increases significantly.

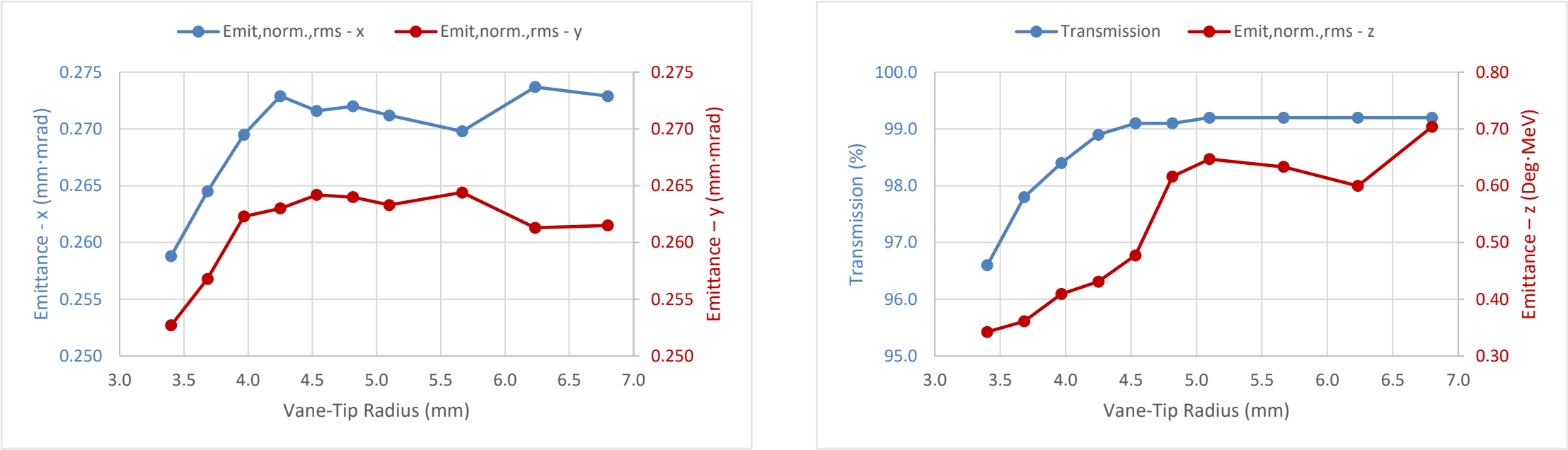

(a) (b)

Fig.6 Evolutions of (a) transverse emittance growth and (b) transmission efficiency and longitudinal emittance growth with different vane-tip radius.

### 3.3 Choice of Vane-Tip Radius

From the above results, the choice of a vane-tip radius of 4.250 mm (or ρ = 0.75) is a good compromise among the considerations for a lower risk of voltage breakdown, smaller emittance growth in transverse and longitudinal phase space, and higher transmission efficiency.

Similarly, the remaining parameters can be optimized. The final RF parameters of the 2D RFQ profile simulated using the CST MWS are presented in Table 3.

Table 3 RF parameters of 2D Proton RFQ.

| Parameters | Proton RFQ |
|---|---|
| Operating frequency (MHz) | 175.998 |
| Quality factor | 16363 |
| Nominal inter-vane voltage (kV) | 80 |
| Power loss per unit length (W/cm) | 257.9 |
| Nearest dipole mode frequency (MHz) | 170.495 |
| Mode separation (MHz) | 5.503 |
| $L_{max}$ (mm) | 164.974 |

## 4. Principle of Pi-Mode Stabilizer Loops

The concept of PISLs is based on mode stabilization by the strong magnetic coupling between two adjacent RFQ quadrants by means of closed-loop couplers, as illustrated in Figure 7. For the quadrupole mode, the total magnetic flux that passes through a surface enclosed by a conducting loop is zero; therefore, PISLs will have negligible effect on the quadrupole field patterns. However, for the dipole modes, the net magnetic flux is nonzero. Then, the dipole modes will be perturbed more, and the induced current detours around the loop will increase the resonant frequency of the lowest dipole mode higher than that of the accelerating quadrupole mode.

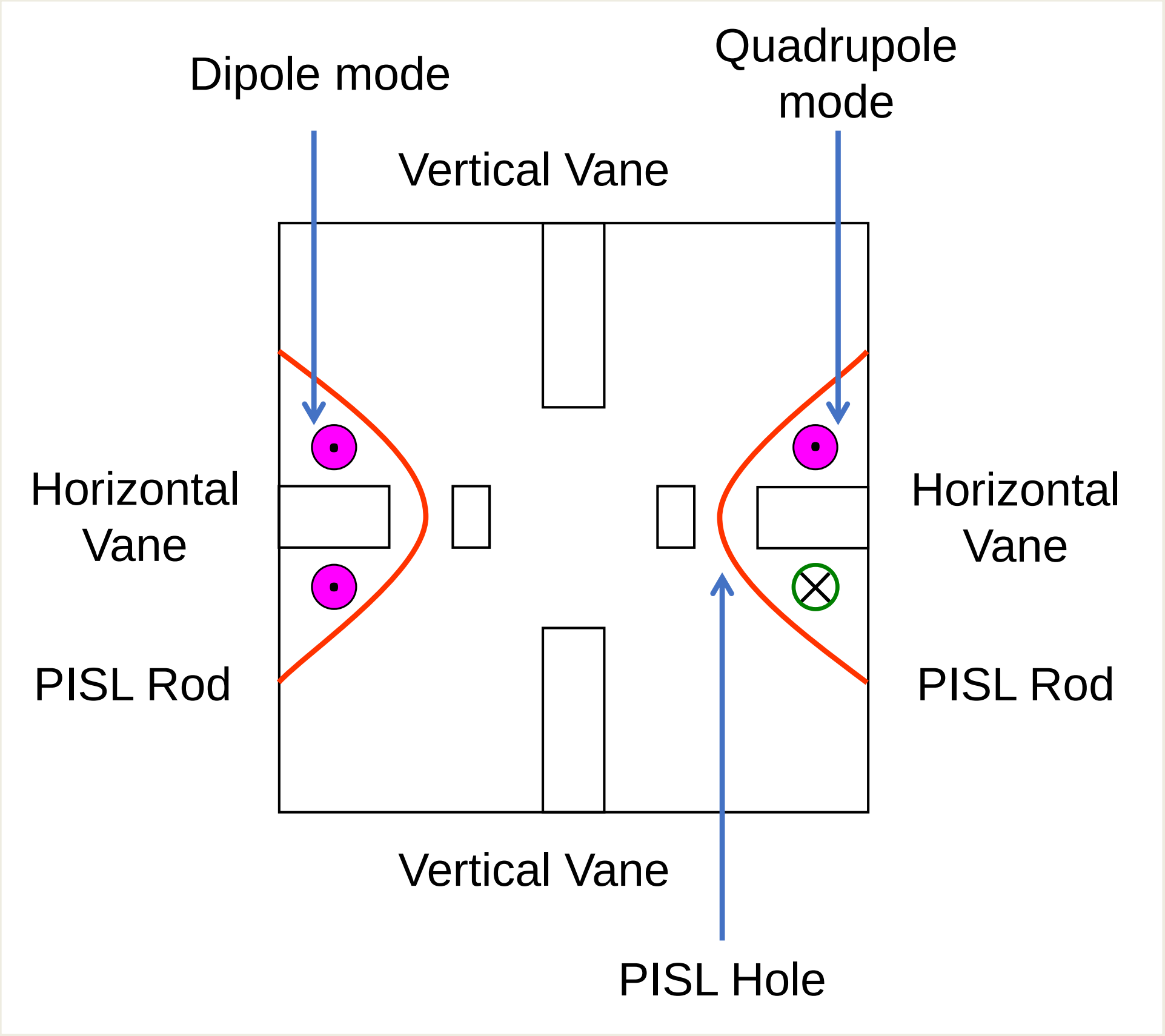

Fig.7 Principle of pi-mode stabilizer loops in a 4-vane RFQ: two curved red lines represent two PISL rods installed in the vertical direction for shifting $TE110^{+}$ mode; two holes on the horizontal vanes are opened for the PISL rods' free access to two RFQ quadrants; shaded circles in magenta with black points used for the magnetic field pointing outward, and green circle with a cross showing the magnetic field pointing inward.

A dedicated arrangement of PISLs must be considered to effectively shift the dipole modes along the RFQ cavity. Figure 8 illustrates the field patterns and associated boundary conditions applied to calculate the degenerate dipole modes ($TE110^+$, $TE110^-$) and quadrupole mode (TE210) in a 4-vane RFQ.

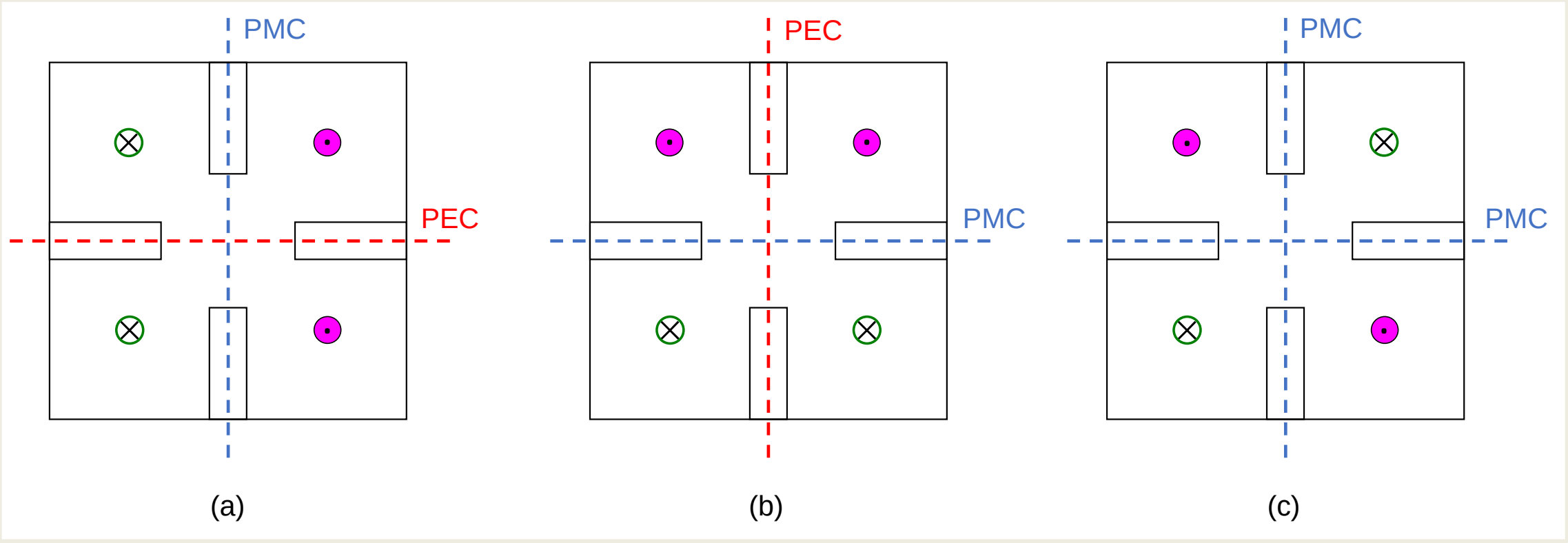

Fig.8 Magnetic field patterns and boundary conditions for calculations of RFQ modes of (a) $TE110^+$, (b) $TE110^-$, and (c) TE210.

If the PISL rods are arranged vertically, as illustrated in Figure 7, the $TE110^+$ mode ($D_0^+$) can be easily shifted. Likewise, the $TE110^-$ mode ($D_0^-$) can be perturbed by the horizontal arrangement of the PISL rods. Thus, pairs of PISL rods alternately installed in the vertical and horizontal directions can noticeably shift the degenerate dipole modes away from the working mode.

### 4.1 Design Considerations of Pi-Mode Stabilizer Loops

It is important to keep beam losses as low as possible in the high intensity RFQs, allowing for reliable and safe maintenance of the machine, which implies that the inter-vane voltage variations (both the quadrupole, $U_q$, and dipole, $U_d$, perturbative components) for the design values should typically stay within a few percent (e.g., < 2%). This aspect is more challenging in a long 4-vane RFQ, whose vane length is several times longer than the RF free-space wavelength, because the machining and misalignment errors deviating from the reference RFQ geometry will cause a mode-mixing challenge between the working TE210 mode and its neighboring dipole TE110 mode, which can deflect the beam transversely and affect beam transmission.

To address this challenge, PISLs are introduced to offer a large mode separation and high field stability against machining and misalignment errors. The design considerations of the PISL optimizations are as follows:

(1) Requirement of mode separation of PISLs ← Limited by the required dipole perturbative component, machine error, and assembly misalignment

(2) Number of PISLs and their mode separation ← Depending on the coupling strength to the dipole modes

(3) Optimization of geometrical parameters of PISLs ← Satisfying the requirement of mode separation and optimizing the power loss of PISL rods

### 4.2 Requirement of Mode Separation of PISLs

A 4-vane RFQ can be described based on a simple 4-wire transmission line model, as illustrated in Figure 9 [30]. $L_i$ and $C_i$ (i = 1, 2, 3, 4) are the inductance per unit length associated with the magnetic field flux through the RFQ quadrant and the inter-vane capacitance per unit length, respectively. $C_a$ and $C_b$ are the capacitance per unit length between two opposite vanes. From transmission line theory and first-order perturbation analysis, the relative amplitude of the dipole perturbative component can be written as follows:

$$\frac{U_{d1n}}{U_{q0}} = \sum_{n=0}^{\infty} \frac{-\sqrt{2}\omega_0^2}{4(\omega_0^2 - \omega_{dn}^2)} \int_0^l \left(\sqrt{\frac{2}{l}} cos(n\pi z)\right)^2 \left(\frac{\Delta C_{Qd1}}{C} + \frac{\Delta L_{Qd1}}{L}\right) dz \ (2)$$

with

$$\Delta C_{Qd1} = \frac{\sqrt{2}(\Delta C_1 - \Delta C_3)}{1+h} (3)$$

$$\Delta L_{Qd1} = \sqrt{2}(\Delta L_1 - \Delta L_3) (4)$$

$$f_{d0} = f_{q0} / \sqrt{1+h} (5)$$

where $\omega_0 = 2\pi \cdot f_{q0}$ is the angular frequency of the operating quadrupole mode, $\omega_{d0} = 2\pi \cdot f_{d0}$ is the angular frequency of the lowest dipole mode (n = 0), $\Delta C_{Qd1}$ and $\Delta L_{Qd1}$ are the capacitance and inductance perturbations caused by the geometrical errors, respectively, and parameter h represents the coupling in an RFQ quadrant.

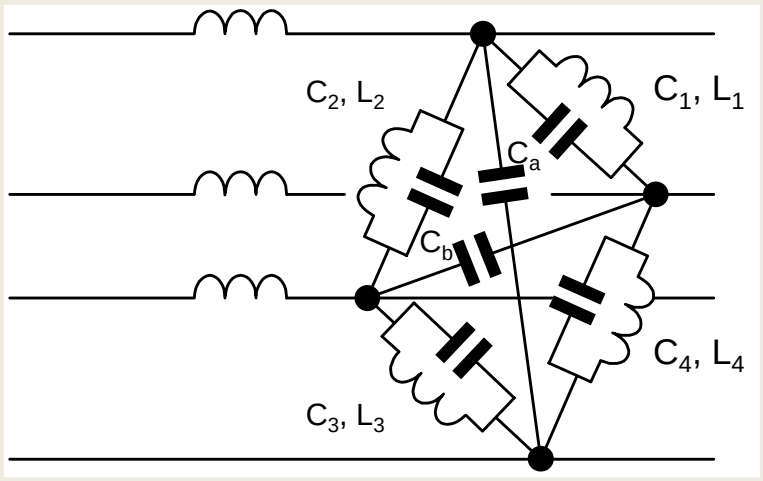

Fig. 9 A simple lumped-circuit model of a 4-vane RFQ.

In an ideal RFQ, we have $C_i$ = C and $L_i$ = L by defining the cutoff frequency $\omega_0 = 1 / \sqrt{LC}$. For further analysis of Eqs. (3) and (4), we can consider the slice quadrant of a 4-vane RFQ structure. Using the error transfer formula, we obtain Eq. (6),

$$\frac{\Delta f}{f_0} = \frac{1}{2}\left(\frac{\Delta C}{C} + \frac{\Delta L}{L}\right) (6)$$

Obviously, ΔC and ΔL depend on the geometric errors of the RFQ profile, which induce a variation in the local cutoff frequency of $f_0$ for the working quadrupole mode

$$\frac{\Delta f}{f_0} = \frac{\left|\chi_{r_0}\right| \cdot \Delta r_0 + \left|\chi_{r_t}\right| \cdot \Delta r_t + \left|\chi_{L_{max}}\right| \cdot \Delta L_{max}}{f_0} (7)$$

In particular, the shunt capacitance, C, depends mainly on the average aperture radius ($r_0$) and vane-tip radius ($r_t$), and the shunt inductance L mainly relies on the half inner width ($L_{max}$) of a 4-vane RFQ cavity. From Eqs. (6) and (7), we can write:

$$\frac{\Delta C}{C} = 2 \cdot \frac{\left|\chi_{r_0}\right| \cdot \Delta r_0 + \left|\chi_{r_t}\right| \cdot \Delta r_t}{f_0} (8)$$

$$\frac{\Delta L}{L} = 2 \cdot \frac{\left|\chi_{L_{max}}\right| \cdot \Delta L_{max}}{f_0} (9)$$

It should be noted that $\chi_{r_t}$ solely depends on the machining error of the vane-tips, and $\chi_{r_0}$ is mainly towing to electrode positioning errors caused by alignment errors and the brazing process. The sensitivity coefficients obtained via 2D RFQ simulations are presented in Table 4.

Table 4 Frequency sensitivity due to geometrical deviations.

| Parameter | Value |
|---|---|
| $\chi_{r_0}$ (MHz/mm) | 9.689 |
| $\chi_{r_t}$ (MHz/mm) | -7.118 |
| $\chi_{L_{max}}$ (MHz/mm) | -1.221 |

Regarding the inductive term, its frequency tuning sensitivity, $\chi_{L_{max}}$, turns out to be approximately one order of magnitude lower than the corresponding capacitive sensitivity. This indicates that the machining of the vane-tips and vane assembly should be highly precise.

The relatively conservative estimates of $\Delta C_{Qd1}$ and $\Delta L_{Qd1}$ in Eqs. (3) and (4) are based on $\Delta C_1 = -\Delta C_3$ and $\delta L_1 = -\Delta L_3$, which can be expressed as

$$\frac{\Delta C_{Qd1}}{C} = \frac{\sqrt{2}(\Delta C_1 - \Delta C_3)}{(1+h)\cdot C} = \frac{2\sqrt{2}}{(1+h)}\frac{\Delta C}{C} \quad (10)$$

$$\frac{\Delta L_{Qd1}}{L} = \frac{\sqrt{2}(\Delta L_1 - \Delta L_3)}{L} = \frac{2\sqrt{2}\Delta L}{L} \quad (11)$$

Therefore, we can study the dependence of the machining error, assembly misalignment, dipole perturbative component, and mode separation requirement. Parameters $\chi_{r_0}$ and $\chi_{r_t}$ are assumed to vary in the ranges of $\Delta r_0$ = 0, ±0.010, ±0.020, …, ±0.100 mm and $\Delta r_t$ = 0, ±0.010, ±0.020, …, ±0.100 mm, respectively, while $\Delta L_{max}$ is assumed to be fixed at 0.030 mm. The values of the required mode separation with dipole perturbative component limited within 2% are illustrated in Figure 10.

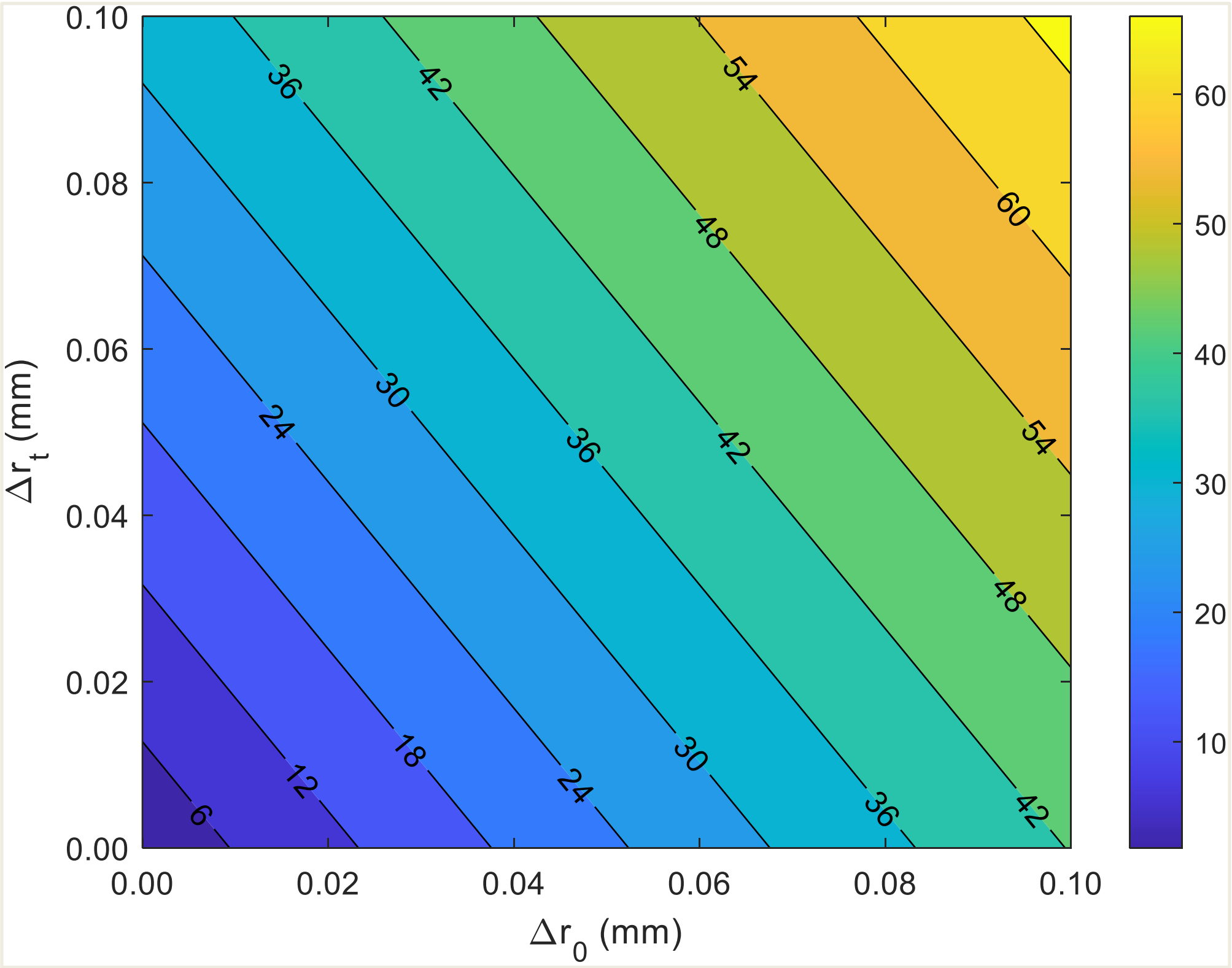

Fig.10 Dependence of the requirement of mode separation (MHz, between operating mode and the nearest dipole mode) for vane-tip machining error, assemble error, and dipole perturbative component within 2%, and $\Delta L_{max}$ is supposed to be 0.030 mm. There is a mathematical technique when handling Eq. (6), which is in a differential form. Because we do not know the plus or minus sign of the errors unless the cavity is made, we have to take the absolute values of Eq. (6), and analyze the cases in Quadrant 1.

It can be inferred that a smaller mode separation will require a very high accuracy of machining and assembly to constrain the dipole perturbative component within 2%. By utilizing PISLs, a large mode separation between the operating quadrupole mode and dangerous dipoles can be obtained to counteract machining and assembly errors. A 4-vane RFQ with a mode separation of 18 MHz could have a maximum acceptable deviation of 0.045 mm for $\Delta r_t$, and 0.035 mm for $\Delta r_0$, thus ensuring high RF stability.

### 4.3 Design Period of Pi-Mode Stabilizer Loops

PISLs are utilized to achieve a desired frequency shift of 18.0 MHz for the nearest dipole modes from the fundamental quadrupole mode. The coupling strength of the dipole modes is strongly influenced by the interval between neighboring PISLs, i.e., the number of PISLs per unit length. Therefore, it is possible to increase the coupling strength by applying more PISLs. One design period of the PISLs is simulated to study its influence on RF properties. As illustrated in Figure 11, the design period of the PISLs is defined by four geometrical parameters, and $L_{max}$ is used only for frequency restoration. A contour plot of the magnetic field patterns of the working mode is also presented.

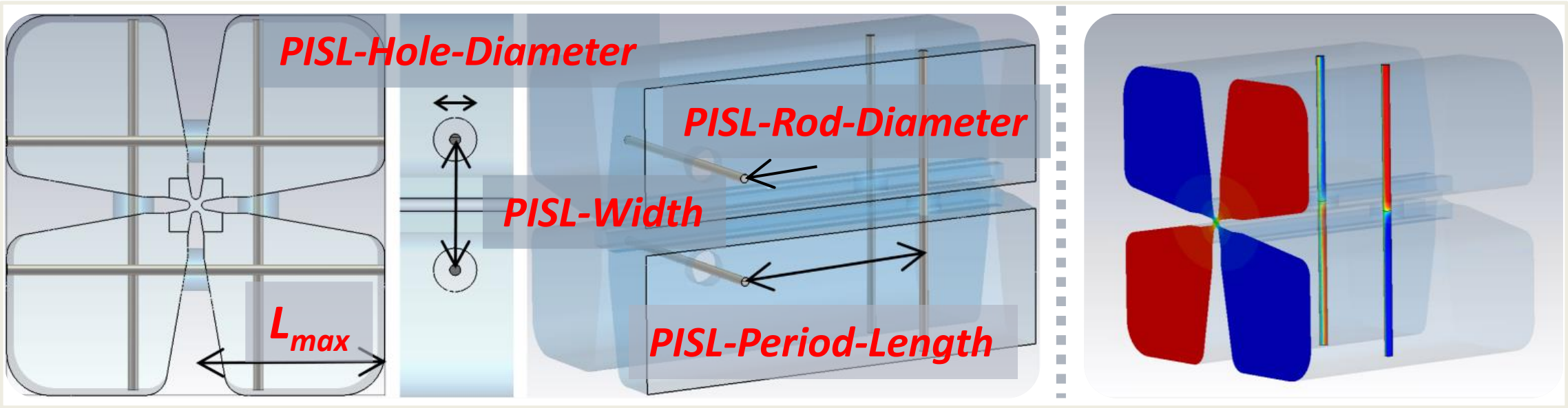

Fig.11 (left) Geometrical parameters of a PISLs period; (right) magnetic field patterns of the working quadrupole mode in one PISLs period, whose color difference indicates that directions of the magnetic field in adjacent quadrants are opposite.

Figure 12 illustrates the mode separation between the unwanted dipole and operating quadrupole modes as a function of the number of pairs of PISL rods per module with a length of 1032 mm. A desired mode separation of 18 MHz can be achieved with the help of four pairs of PISL rods per module, which is adequate to suppress the dipole perturbative component below 2%. Although a mode separation larger than 18 MHz can be realized by more pairs (> 4) of PISL rods, the cavity performance degrades at the cost of the quality factor. Therefore, the decision of four pairs of PISL rods per RFQ module (or a PISL period length of 258 mm) is more favorable for achieving high RF stability and good cavity performance.

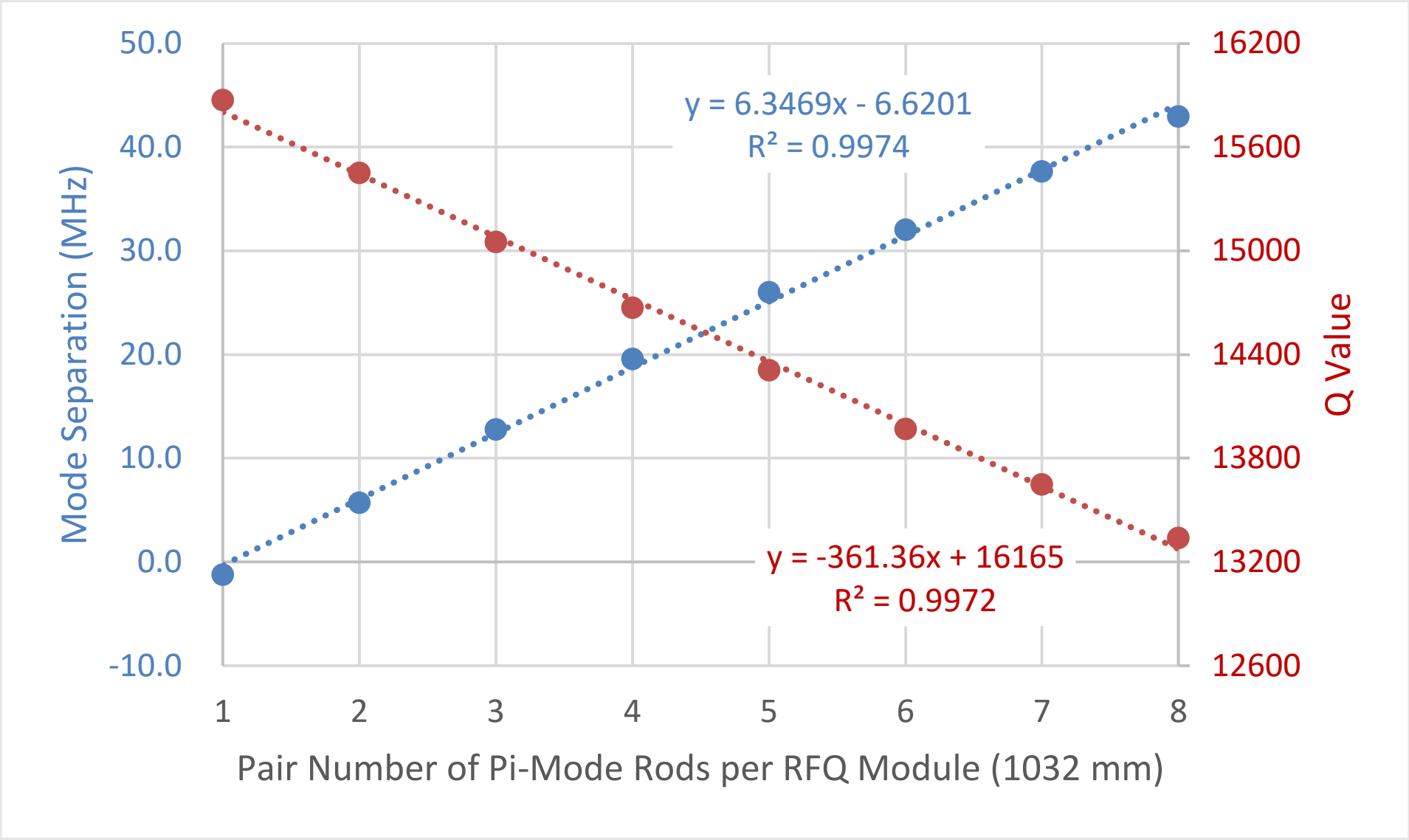

Fig.12 Mode separation and quality factor versus pairs of PISL rods per RFQ module.

The remaining geometrical parameters are optimized in a similar manner. It should be noted that a smaller PISL width (116 mm) is preferred owing to a larger mode separation, as can be observed from Figure 13, but the intersection between the PISL rod and vane corner should be avoided.

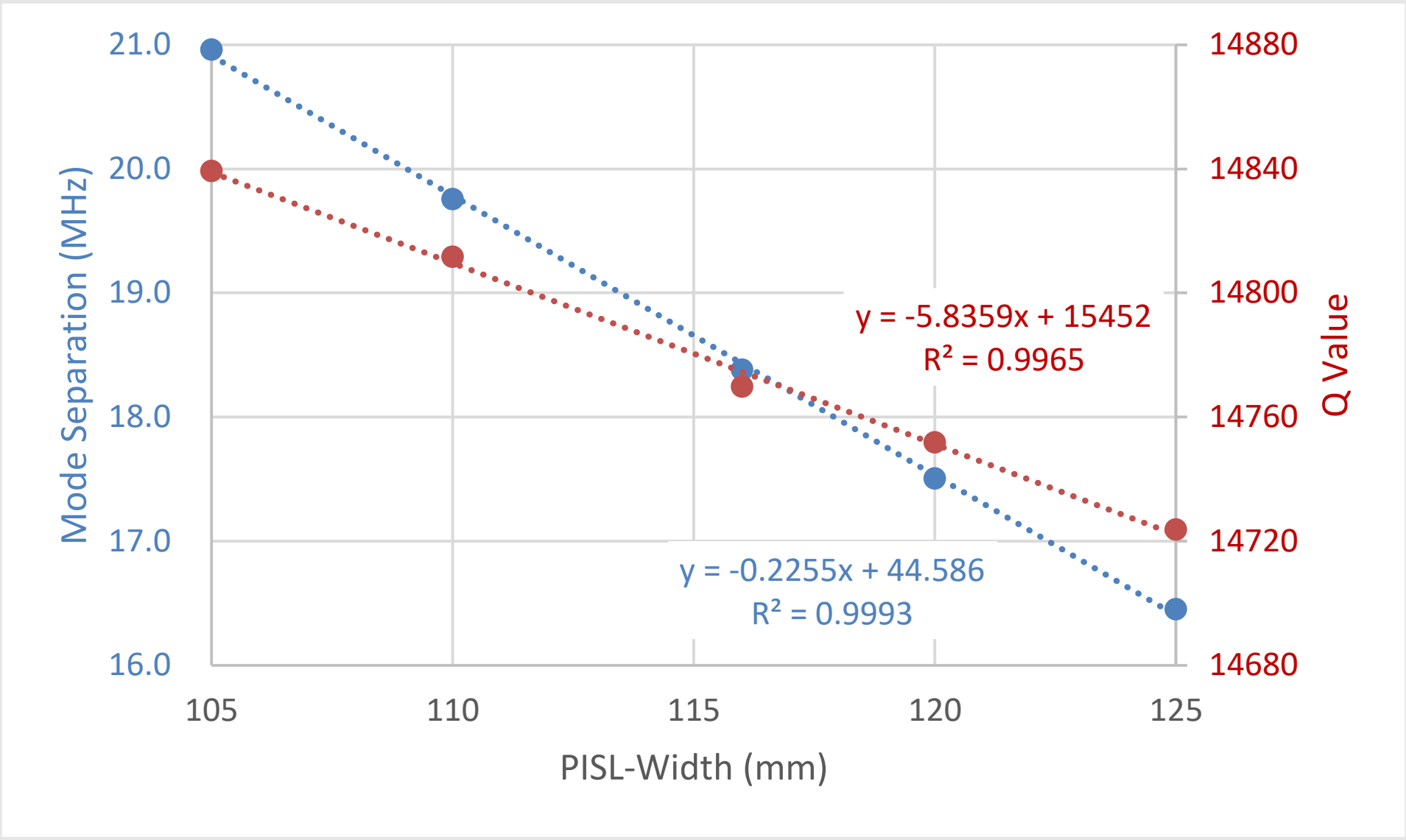

Fig.13 Mode separation and quality factor versus different PISL-Width.

In addition, the diameters of the PISL hole and rod are mainly related to controlling the power loss on the PISL rods, but tuning the PISL hole diameter is the most effective approach (refer to Figure 14 (a)). The PISL hole diameter is determined to be 50 mm for a modest power loss on the PISL rod. The diameter of the PISL rod is mainly constrained by its cooling capacity. Based on the design experience in Refs. [22,23,25], a PISL rod with an outer diameter of 10 mm and inner diameter of 5 mm is able to remove an average (10 % duty cycle) heat load of approximately 32 W in this design.

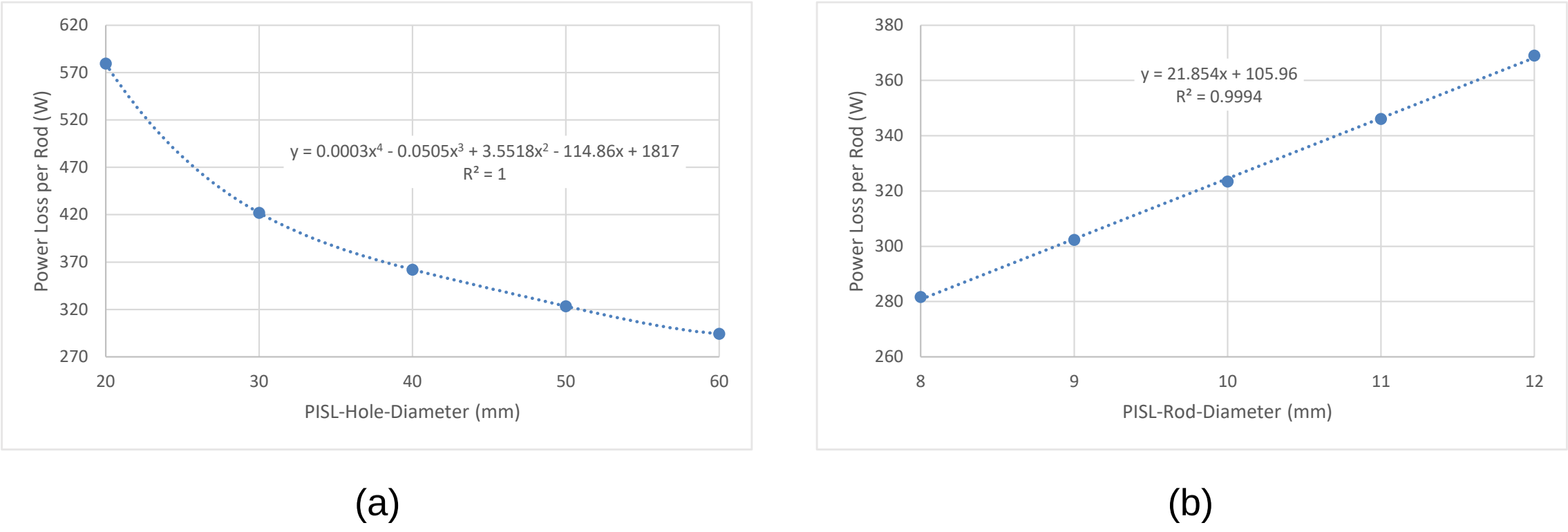

Fig.14 Power loss per rod versus different values of (a) PISL-Hole-Diameter and (b) PISL-Rod-Diameter.

The RF parameters simulated with CST MWS after the PISL period optimization are presented in Table 5. Without PISLs, the mode separation is only 5.503 MHz, as opposed to 18.160 MHz with PISLs.

Table 5 RF simulation results of a Pi-mode stabilizer period.

| Parameter | Value |
|---|---|
| Quadrupole mode frequency (MHz) | 175.996 |
| Quality factor | 14829 |
| Nominal inter-vane voltage (kV) | 80 |
| Power loss per rod (W) | 323.4 |
| Nearest dipole mode frequency (MHz) | 194.156 |
| Mode separation (MHz) | 18.160 |
| $L_{max}$ (mm) | 160.738 |

## 5. Design of Tuner Period

The tuner period is designed to provide a large tuning range of approximately ±1.0% of the center frequency. The RFQ cavity is equipped with 100 slug tuners and equally distributed in the four quadrants. The simulation results of one tuner period presented in Table 6, are illustrated in Figure 15. One tuner period is 207.5 mm in length along the longitudinal direction, and each tuner is 60 mm in diameter with 20 mm nominal insertion depth, giving a tuning range of -1.740 MHz to 2.090 MHz. The tuning sensitivity per tuner near its nominal insertion depth is 24.8 kHz/mm using a curve fitting technique (refer to Figure 16).

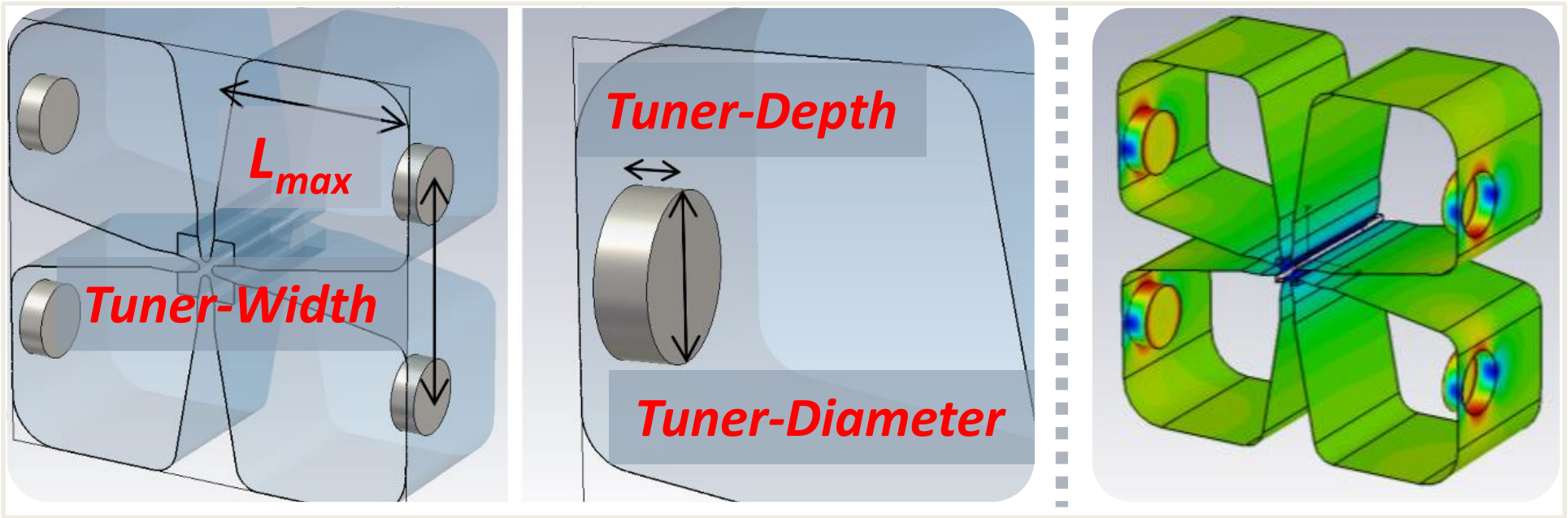

Fig.15 (left) Geometrical parameters of a tuner period; (right) contour plot of surface current density of the working quadrupole mode on the resonator wall.

Table 6 RF simulation results of a tuner period.

| Parameter | Value |
|---|---|
| Quadrupole mode frequency (MHz) | 175.995 |
| Quality factor | 15545 |
| Nominal inter-vane voltage (kV) | 80 |
| Nominal insertion depth (mm) | 20 |
| Power loss per tuner (W) | 123.5 |
| Tuning sensitivity (kHz/mm) | 24.8 |
| $L_{max}$ (mm) | 160.738 |

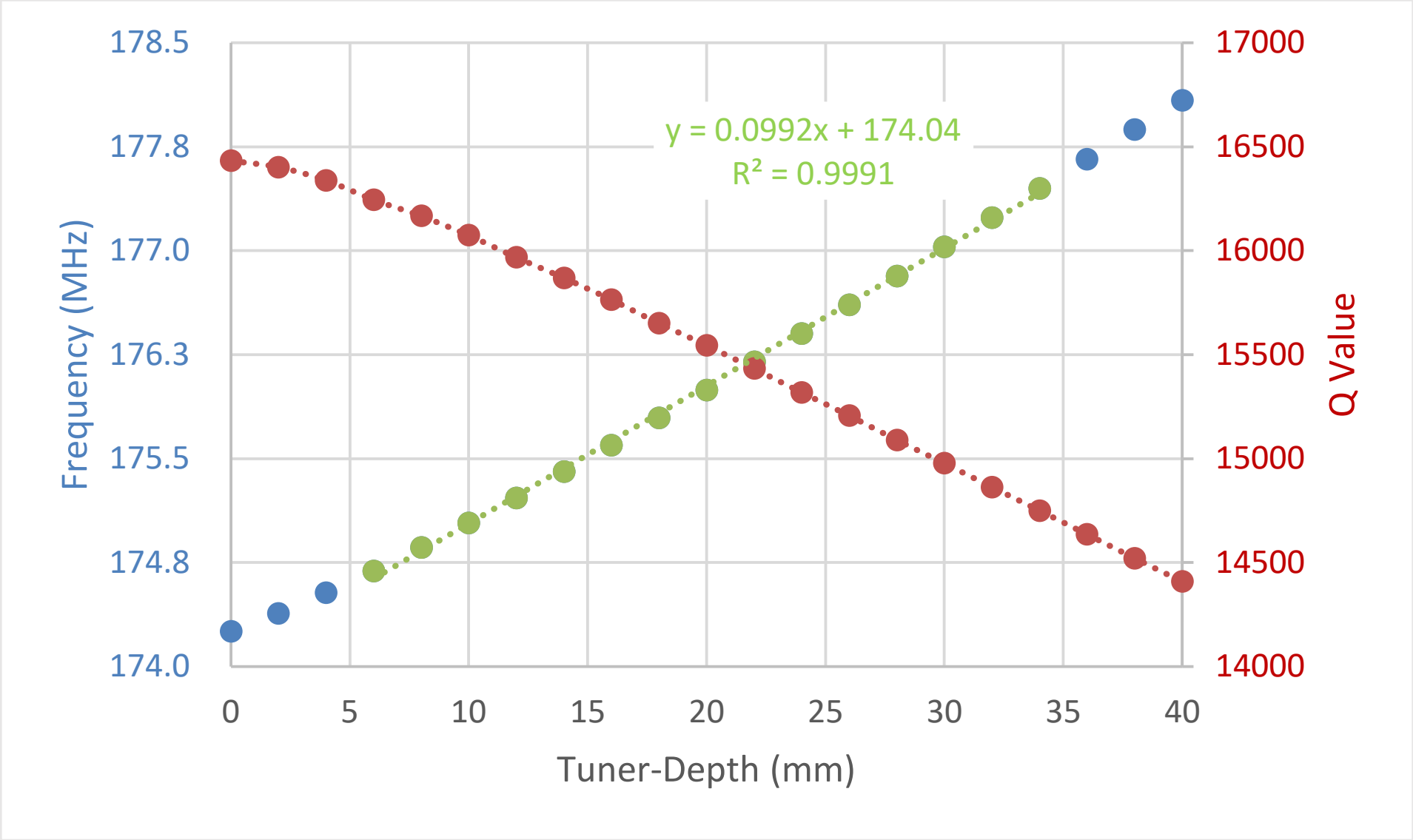

Fig.16 Frequency response (colors in blue and green) and quality factor (color in red) versus different Tuner-Depth, and the tuning region marked by green has a good linearity and is considered to be a favorable tuning region.

Because the maximum local frequency shift due to the modulation is -219.8 kHz within a unit cell length model with a modulation factor of 2.382 (Cell #262), it can be compensated by using plug tuners with sufficient margins. Consequently, it is not an issue for the design to use unmodulated vane-tips to save time and memory.

## 6. Cavity Tuning with Perfect Magnetic Boundary Conditions

The spacing between the slug tuners and PISL rods is not regular. To consider such a 3D effect, a full-length cavity model installed with complete sets of 20 pairs of PISL rods and 100 tuners is put into the simulations, with cavity ends that are set to perfect magnetic boundary conditions. With this model, the final tuning of the RFQ main body is performed, and the tuning parameter, $L_{max}$ = 162.300 mm, is fixed.

## 7. Design of End-Cells of RFQ Cavity

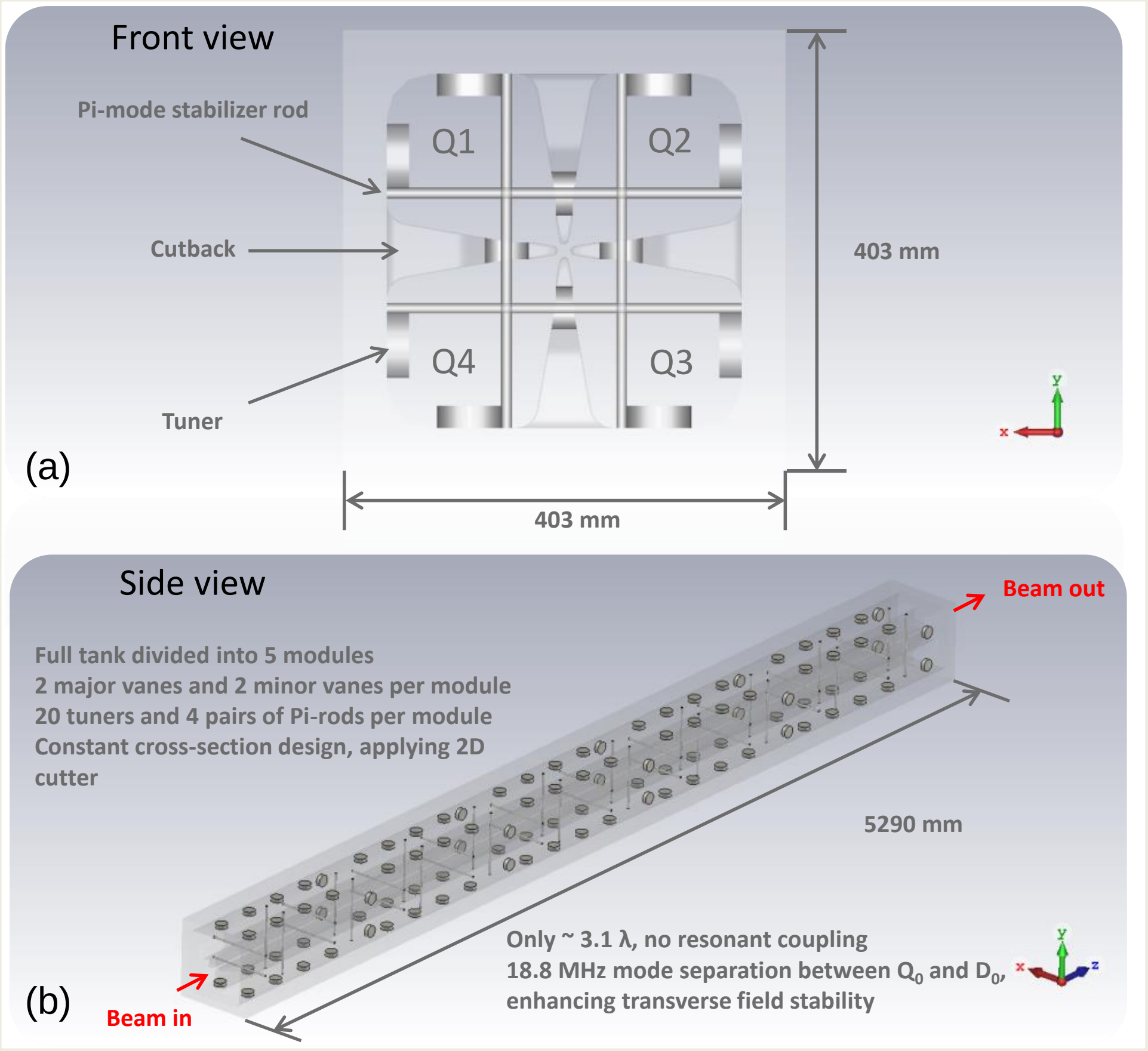

Fig.17 Schematic view of 176 MHz RFQ: (a) front view, (b) side view.

A schematic view of the 176 MHz RFQ with end-cells is illustrated in Figure 17. When the perfect RFQ in Section 6 is closed with end plates, the electric boundary conditions force the gap electric field to reach zero, resulting in a TE211 mode with a half wavelength variation along the z-direction. To excite a pure TE210 mode, the end-cells of the RFQ cavity must be modified by opening the cutbacks, which will help circulate the magnetic fields in the neighboring quadrants, thus resulting in a flat transverse field distribution along the RFQ resonator. The basic geometry of the cutback is illustrated in Fig. 18. However, the cutback depth ($D_{in}$ or $D_{out}$) is the most influential parameter for the field flatness. The main effect of the parameter H is to determine the end-cell capacitances in the presence of the radial matching section (RMS) gap and fringe field section (FFS) gap, which are given by the ParmteqM code. In our case, H is 45 mm, and the cutback angle α is assumed to be 60°. After a series of dedicated simulations, the input and output cutback depths are optimized to 77.5 mm and 73.0 mm, respectively. The electric field distribution between vane-tips is flat well within ±1%.

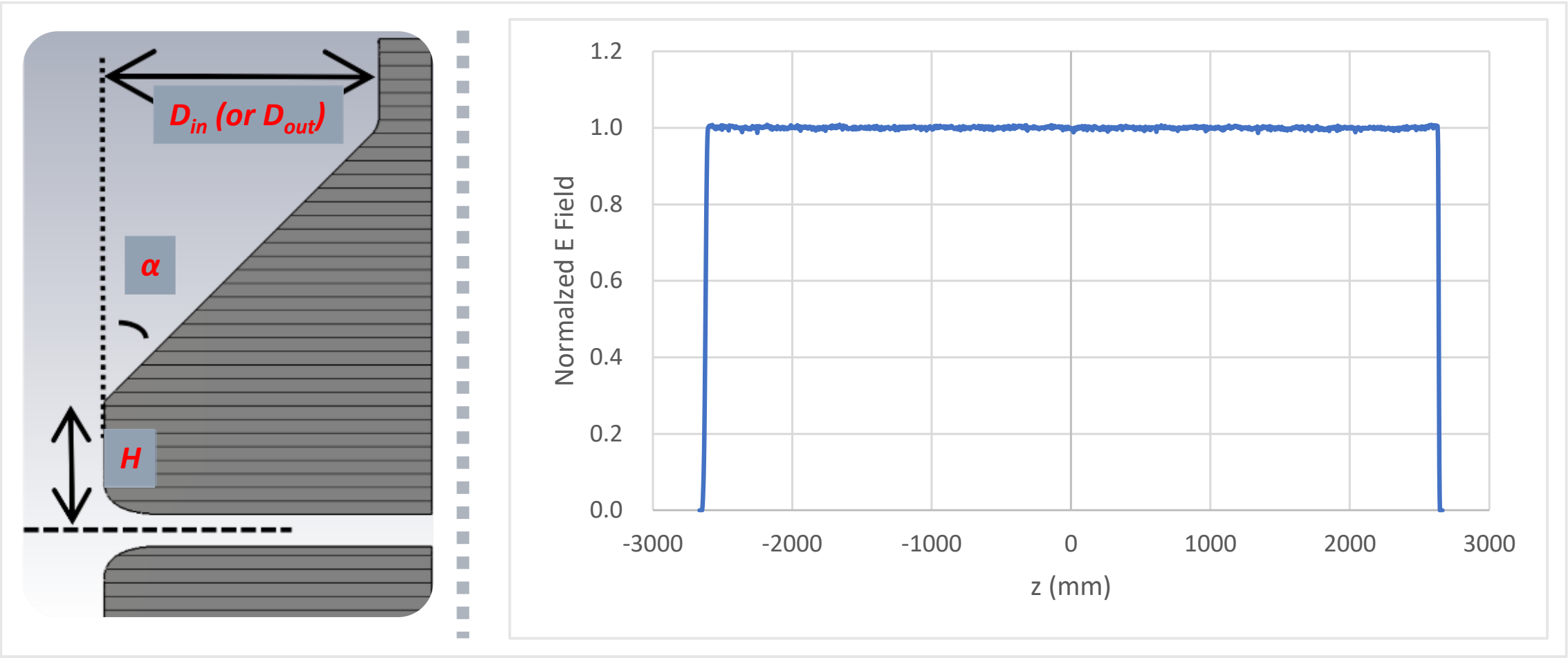

Fig.18 (left) Schematic diagram of RFQ cutbacks with triangle shape; (right) normalized electric field distribution along RFQ accelerator after dedicated tuning of cutback depths (field flatness within ±1.0%).

Table 6 presents the power dissipation values for separate parts of the RFQ with an electrical conductivity of $5.8 \times 10^7$ S/m. Depending on the quality of the copper and brazing, the real total power loss could be increased by 20%.

Table 6 Power dissipations on separate parts simulated in a full-length cavity.

| Part | Value (kW) | % |
|---|---|---|
| Cavity wall including vanes | 139.469 | 79.46 |
| Input cutbacks, 4 units | 3.675 | 2.09 |
| Output cutbacks, 4 units | 4.010 | 2.28 |
| PISL rods, 20 pairs | 13.230 | 7.54 |
| Tuners, 100 units | 14.258 | 8.12 |
| Front-end plate | 0.345 | 0.20 |
| Back-end plate | 0.542 | 0.31 |
| Total power loss (PEC) | 175.530 | 100 |
| Total power loss (with 20% Margin) | 210.636 | / |

Considering a 10% duty factor and 1.2 scaling factor for a real loss estimation, the maximum power loss density at the input and output cutbacks are 6.56 W/cm$^2$ and 7.03 W/cm$^2$, respectively. Compared to the maximum power loss densities of some CW RFQs in the world [10,14,25,31,32,33], it can be concluded that an average heat loss density of 7.03 W/cm$^2$ is modest, which is beneficial to high duty cycle operation.

## 8. Multipacting Phenomenon

The multipacting phenomenon is a resonant discharge that is often observed in high-power microwave devices and may result in damages to the RF components and distortions of the RF signals. Multipacting analysis is performed using 3D CST Particle Studio code [29] implemented with a particle-in-cell solver. The initial particle source is defined on the inner surface of the copper wall using the Furman-Pivi emission model [34]. The emission pulse is assumed to be a Gaussian beam. The multipactor indicators follow the definitions in Ref. [35],

$$\langle SEY \rangle = \frac{I_{emission}}{I_{collission}} \quad (12)$$

$$W_{collission}(eV) = \frac{P_{collision}(W)}{I_{collision}(A)} \quad (13)$$

where <SEY> is the average secondary emission yield, $I_{collission}$ is the incident secondary electron current, $I_{emission}$ is the electron current emitted from the RF surface, $W_{collision}$ indicates the average collision energy, and $P_{collision}$ is the power of the incident secondary electrons.

The copper power levels used in the simulations are normalized to the operation level. A series of simulations with a normalized power factor ($P/P_{norm}$) ranging from 0.01 – 1.21 with a step of 0.04, are studied.

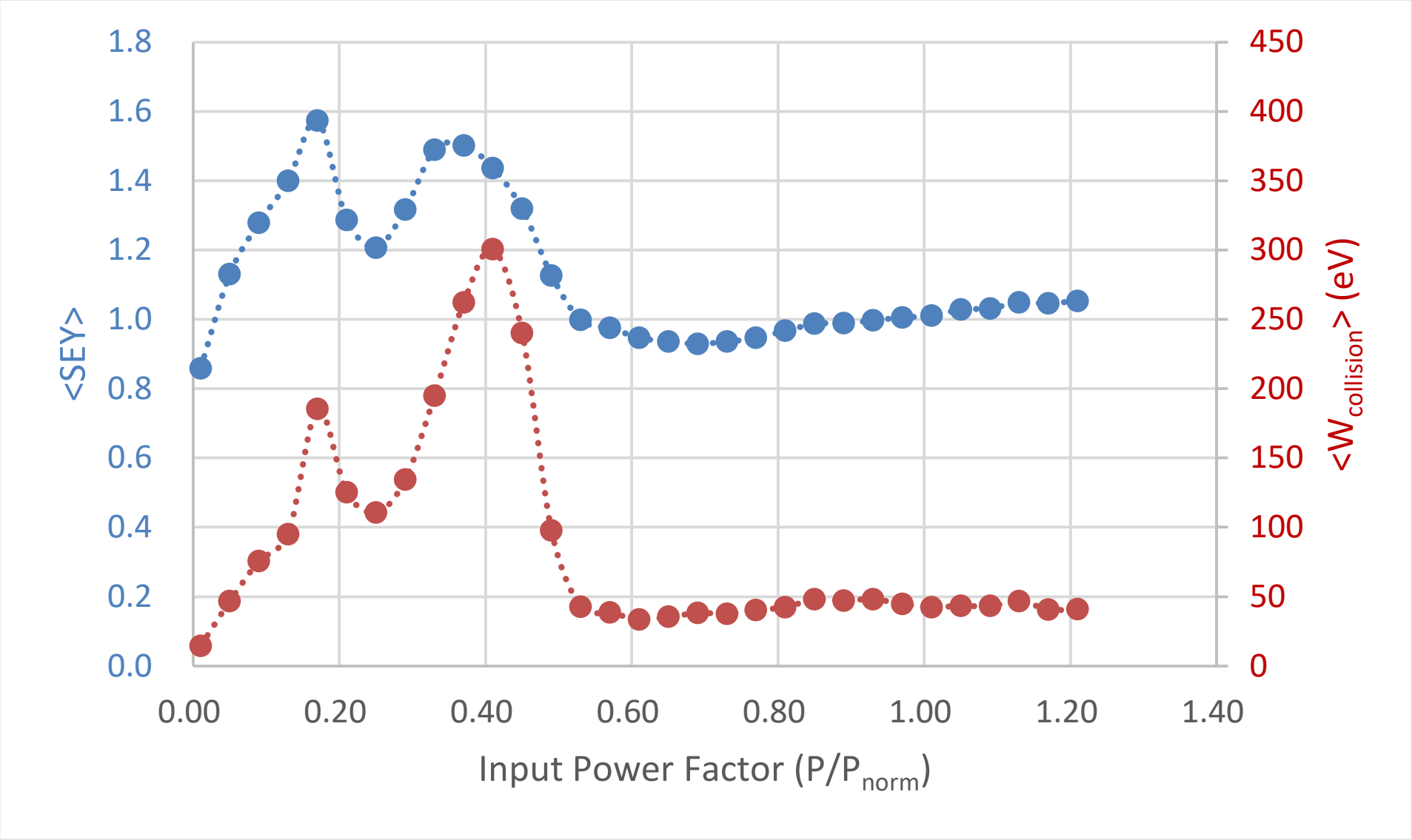

Fig. 19 Simulated multipacting results at different power factors.

The particle-tracking results are illustrated in Figure 19. Two significant multipacting barriers covering a range of $P/P_{norm}$ = 0.09 – 0.53, have been determined. They are intense resonant multipacting of 1 – 2 orders, which means that a long processing time is required to overcome conditioning difficulties. After passing through two multipacting peaks, the average collision energy of the secondaries drops to approximately 50 eV, and the surface cleanliness is expected to improve. Thus, the risk of multipacting at the designed power level is low.

## 9. Conclusion

The physics design results and high-frequency structure optimizations of a high-duty factor 4-vane RFQ with pi-mode stabilizer loops (PISLs) were presented. A systematic, quantitative model was developed to analyze the RF stability of a 4-vane RFQ with PISLs and the theoretical error tolerance ranges. A complete RFQ cavity with 100 tuners, 20 pairs of PISL rods, and two end-cells was designed and simulated. Multipacting simulations were also performed, and a low risk of multipacting near the operating power level was revealed. The cavity power was evaluated to be 211 kW at a nominal vane voltage of 80 kV, and this RFQ delivered an acceleration of 2.5 MeV to an 80 mA proton beam within 5.3 m.